\documentclass[twocolumn,aps,showpacs,pre,amsmath]{revtex4-1}

\usepackage{amssymb}
\usepackage{graphicx}
\usepackage{dsfont}

\begin{document}

\title{Aharonov-Bohm interferences in polycrystalline graphene}

\author{V. Hung Nguyen\footnote{E-mail: viet-hung.nguyen@uclouvain.be} and J.-C. Charlier} \address{Institute of Condensed Matter and Nanosciences, Universit\'{e} catholique de Louvain (UCLouvain), Chemin des \'{e}toiles 8, B-1348 Louvain-la-Neuve, Belgium}

\begin{abstract}
	Aharonov-Bohm (AB) interferences in the quantum Hall regime can be achieved, provided that electrons are able to transmit between two edge channels in nanostructures. Pioneering approaches include quantum point contacts in 2DEG systems, bipolar graphene \textit{p-n} junctions, and magnetic field heterostructures. In this work, defect scattering is proposed as an alternative mechanism to achieve AB interferences in polycrystalline graphene. Indeed, due to such scattering, the extended defects across the sample can act as  tunneling paths connecting quantum Hall edge channels. Consequently, strong AB oscillations in the conductance are predicted in polycrystalline graphene systems with two parallel grain boundaries. In addition, this general approach is demonstrated to be applicable to nano-systems containing two graphene barriers with functional impurities and perspectively, can also be extended to similar systems of 2D materials beyond graphene.  
\end{abstract}

\maketitle

\section{Introduction}

Graphene and other two-dimensional (2D) layered materials have recently become very attractive in several research fields \cite{Ferrari2015} because of novel physical properties, as a consequence of their atomically-thin 2D crystalline structure. The study of their defects (always present in real systems) is hence highly desirable \cite{Araujoa2012,Chen2012,Terrones2012,Yazyev2014,Cummings2014a,Lin2016,Isacsson2017} in order to estimate their impacts on various properties, but also to possibly explore novel ways to tune the electronic structure of these 2D materials at the nanoscale. While the former is mandatory to fully understand the properties of practical 2D materials, the latter is helpful for enlarging their applications and concurrently exploring new ones. 

In particular, to achieve large scale graphene samples with high structural quality, the chemical vapor deposition (CVD) method is known to be the best technique, although yielding graphene in a polycrystalline form \cite{Ferrari2015,Yazyev2014,Isacsson2017}. This polycrystalline graphene contains pristine grains separated by extended structural defects consisting of a random 1D distribution of non-hexagonal (i.e., pentagon, heptagon, octagon, etc.) rings, vacancies and impurities. In general, these defects represent an important scattering source that limits drastically the carrier mobility in CVD graphene samples \cite{Tuan2013} and consequently impedes the performance of graphene-based electronic devices \cite{Cummings2014a,Isacsson2017}. The electronic transport across graphene grain boundaries has been concurrently demonstrated to be strongly dependent of their structure, i.e., being either highly transparent or perfectly reflective over significant energy ranges \cite{Yazyev2010,Dechamps2018} and tunable by strains \cite{Kumar2012,Nguyen2016a}. However, graphene grain boundaries have also been suggested to be usable to explore novel phenomena such as valley filtering \cite{Gunlycke2010,Nguyen2016b}, creating one-dimensional (1D) topologically conducting states \cite{Lahiri2010,Cma2014}, for electron optics \cite{Nguyen2016b}, and for electronic waveguides \cite{Mark2012}. In this regard, achieving ordered structures of defects is required and this practical challenge has also been experimentally realized \cite{Chen2014,Yang2014}.

The Aharonov-Bohm (AB) effect is a fundamental phenomenon of quantum interference related to the transmission of electrons in closed trajectories in 2D systems when a magnetic field (\textit{B}-field) is applied perpendicularly \cite{Aharonov1959}. Besides its fundamental significance, this effect is extremely useful for specific applications in mesoscopic interferometers such as electron Sagnac gyroscopes \cite{Search2009} and quantum computing systems \cite{Brassard1998}. The AB interferences in conventional 2DEG have been actually observed in two types of nanoscale systems: in quantum-ring structures \cite{Levy1990,Mailly1993} and in systems containing two nano-constrictions in the quantum Hall regime \cite{Chamon1997,Camino2007,Nakamura2019}. In the latter case, quantum Hall edge states are obtained when a strong magnetic field is applied. The interaction between these edge states at the constrictions creates closed trajectories for electrons in the internal cavity, i.e., they can circulate along the cavity edge \cite{Chamon1997}. Such an electron propagation results in an interference effect, i.e., AB oscillations in the conductance. An advantage of this approach is to benefit from the large coherence length of quantum Hall edge states \cite{Roulleau2008}, which is mandatory for electron interferometers.

With its quasi-perfect 2D crystalline structure, graphene is certainly an ideal candidate to design AB interferometers. Actually, besides quantum-ring structures reported in several works (see ref. \cite{Jorg2012} and references therein), graphene AB interferometers in the quantum Hall regime have been also explored, particularly, using graphene \textit{p-n} junctions \cite{Morikawa2015,Kolasinska2016} and \textit{B}-field heterostructures \cite{Nguyen2019}. Although completely different from the interferometers based on nano-constrictions mentioned above, the AB interferences in both these cases are essentially due to the interaction between opposite propagating edge states obtained in different doped / opposite \textit{B}-field zones. Essentially, the interedge interaction induces the tunneling paths connecting quantum Hall edge channels. Thereby, an internal zone, where electrons can circulate along the edge, can also be created in analogy with that obtained in the nano-constriction 2DEG systems. 

Within such a scientific context, the magneto-transport in polycrystalline graphene is investigated in this work, suggesting an alternative approach to achieve AB interferences in 2D layered material systems in the quantum Hall regime.

The main idea was suggested by the fact that in addition to the generated scattering effects, grain boundaries can also be used to create 1D conducting channels across graphene samples. Actually, the creation of these 1D conducting channels has been experimentally demonstrated \cite{Lahiri2010,Cma2014,Adina2016} and explained as a consequence of the "self-doping" effect by localized electronic states around these extended defects. In the presence of extended defects, gate tunable and topologically protected states can also be observed in bilayer graphene domain wall systems \cite{Jasklski2016,Jasklski2018}. Most importantly, magneto-transport in polycrystalline graphene has been investigated \cite{Cummings2014,Bergvall2015,Lago2015,Phillips2017} and these conducting channels are shown to be able to connect quantum Hall edge states, similar to those obtained in graphene \textit{p-n} junctions and \textit{B}-field heterostructures as mentioned above. 
 
On this basis, the goal of this work is to demonstrate that the defect scatterings (consequently, localized electronic states around the grain boundaries) can be exploited as a novel mechanism to achieve AB interferences in the quantum Hall regime, particularly, in polycrystalline graphene systems with two grain boundaries as schematized in Fig.\ref{fig_sim1}. Experimentally, two possibilities to produce such polycrystalline systems have been already demonstrated. On the one hand, synthesized on polycrystalline Cu foils by ambient CVD \cite{Qyu2011} or using the atmospheric pressure CVD technique \cite{Yasaei2014}, polycrystalline graphene systems containing two or a few grains, coupled in series, can be achieved and controlled. On the other hand, some experimental technique for self-catalyzing grown of extended defect lines in graphene have also been developed \cite{Chen2014,Robertson2012}. These techniques clearly demonstrate the possibility of atomically precise engineering of defect lines in graphene. Besides the observation of AB interferences in these structural defective systems, we will additionally demonstrate that the approach presented in this work can also be extended to graphene systems containing two parallel graphenic barriers with functional impurities, particularly, including oxygen impurities, which can be experimentally achieved as in refs. \cite{Xwu2008,Byun2011,Masubuchi2011,Wang2012,Lee2016a,Lee2016b,Dago2018}.

\section{Calculation methodologies}

\begin{figure}[!b]
	\centering
	\includegraphics[width=3.4in]{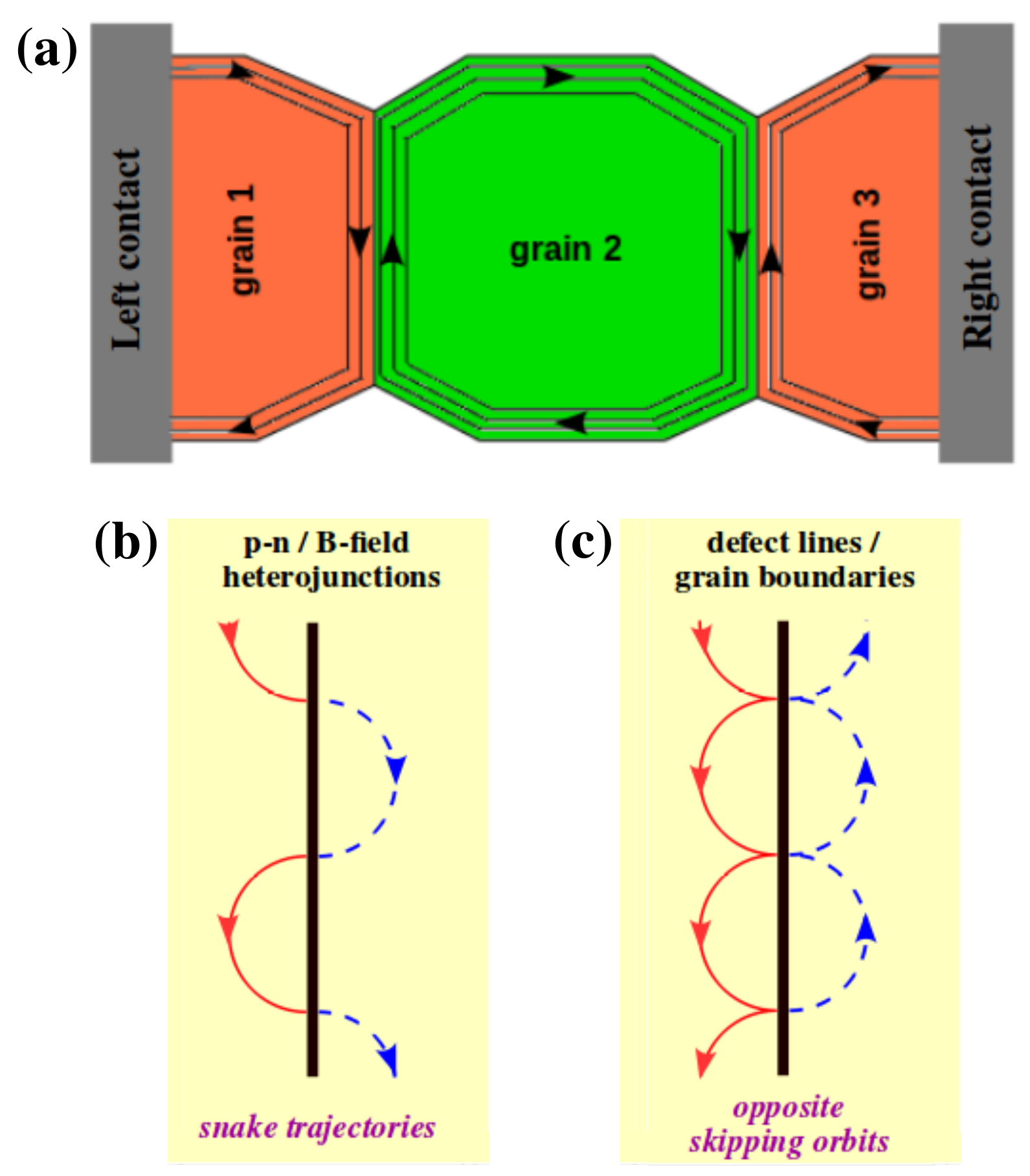}
	\caption{Aharonov-Bohm interferences in polycrystalline graphene in the quantum Hall regime. (a) Electron propagation in three different graphene grains when a high magnetic field is applied. (c) Opposite skipping orbits of electrons around the grain boundaries, compared to (b) snake trajectories for graphene \textit{p-n} junctions and \textit{B-}field heterostructures.}
	\label{fig_sim1}
\end{figure}

In this work, we investigate the magneto-transport through polycrystalline graphene containing three connected-neighboring grains, which can be, as mentioned, experimentally achieved as in refs. \cite{Qyu2011,Yasaei2014}. To mimic their practical properties, irregular edges as illustrated in Fig.\ref{fig_sim1} are considered while graphene grains which generally exhibit different crystallographic orientations and structural disorders along/around the grain boundaries are also taken into account. The presence of non-hexagonal carbon rings can, in principle, induce local deformations (local strains) in graphene lattices in the surrounding zone of grain boundaries. Hence, classical molecular dynamics simulations as in refs. \cite{Dechamps2018,Nguyen2016a} were first performed to relax the atomic structure. The $p_z$ tight-binding Hamiltonian was then employed to compute their electronic properties. In particular, when a magnetic field is perpendicularly applied, this tight-binding Hamiltonian reads:
\begin{equation}
H =  \sum\limits_{\left\langle n,m \right\rangle} t_{nm} e^{i\phi_{nm}} c^\dagger_n c_m
\end{equation}
where $t_{nm}$ corresponds the nearest-neighbor hopping energies, and $\phi_{nm} = \frac{e}{\hbar} \int_{\mathbf{r}_n}^{\mathbf{r}_m} \mathbf{A}(\mathbf{r}) d\mathbf{r}$ is the Peierls phase describing the magnetic field effects. Taking into account the lattice deformation mentioned above, these hoping energies are computed by $t_{nm} = t_0 \exp\left( -3.37 \left\lbrace r_{nm}/r_0 - 1\right\rbrace  \right)$ \cite{Pereira2009} where $t_0 = -2.7$ eV and $r_0 = 0.142$ nm represent the hoping energy and \textit{C-C} bond length of pristine graphene, respectively. In addition, to model the disordered grain boundaries, vacancies are randomly generated around and along the ordered ones.
\begin{figure*}[!t]
	\centering
	\includegraphics[width=7.0in]{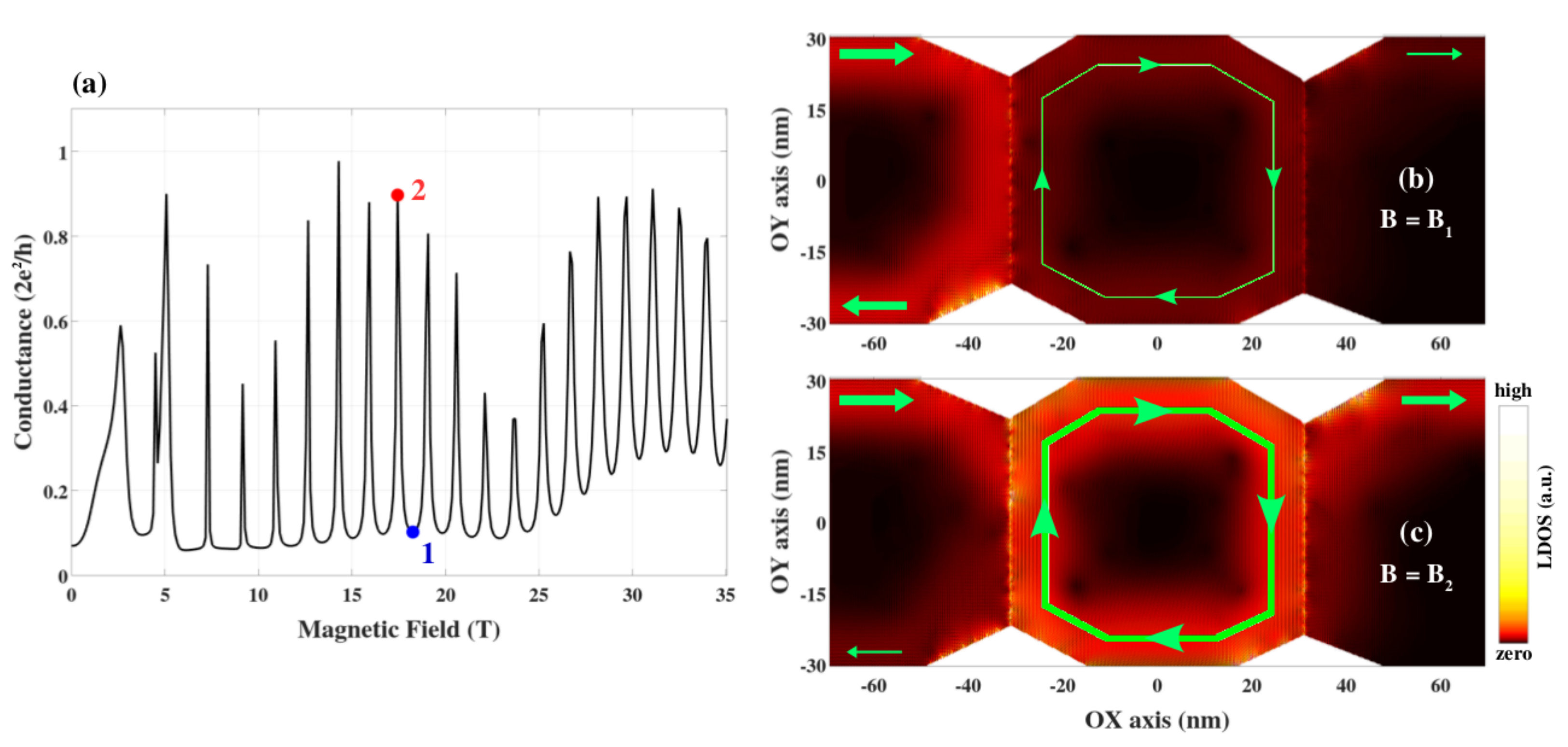}
	\caption{Aharonov-Bohm interferences in polycrystalline graphene. (a) Conductance in the system illustrated in Fig.\ref{fig_sim1} at the Fermi energy $E_F = 70$ meV is plotted as a function of \textit{B-}field. The maximum width (in the Oy axis) of graphene grains is 60 nm while the maximum length of the grain 2 (distance between grains 1 and 3, in the Ox axis) is about 62 nm (see in (b,c)). The atomic structure of the system is described in detail in the Supplementary Information \cite{Nguyen_suppl}. (b,c) Local density of left-injected electronic states (LDOS) at two different \textit{B-}fields $B_1$ and $B_2$ marked in (a), illustrating the electron propagation in the destructive and constructive states, respectively. The superimposed green lines (with arrows) illustrate the propagation trajectories of electrons.}
	\label{fig_sim2}
\end{figure*}

To model polycrystalline graphene systems with oxygen impurities, the Hamiltonian above is adjusted by a fit to Density Functional Theory (DFT) calculations \cite{Nguyen_suppl}. In particular, the impurity effects are effectively computed by introducing an onsite energy $\varepsilon_{on} = 28$ eV to carbon sites, which directly interact with the impurities. This adjusted Hamiltonian is demonstrated to reproduce quite accurately the DFT electronic bandstructure of the system in a low energy range (i.e., $\left| E \right| \lesssim$ 500 meV \cite{Nguyen_suppl}) and with reasonable oxygen impurity densities.

Finally, the tight binding Hamiltonians described above were solved and the transport quantities (transmission probability, conductance, local density of states) were computed using the Green's function technique and the Landauer formalism \cite{Nguyen2013}.

\section{Results and discussions}

\begin{figure*}[!t]
	\centering
	\includegraphics[width=7.0in]{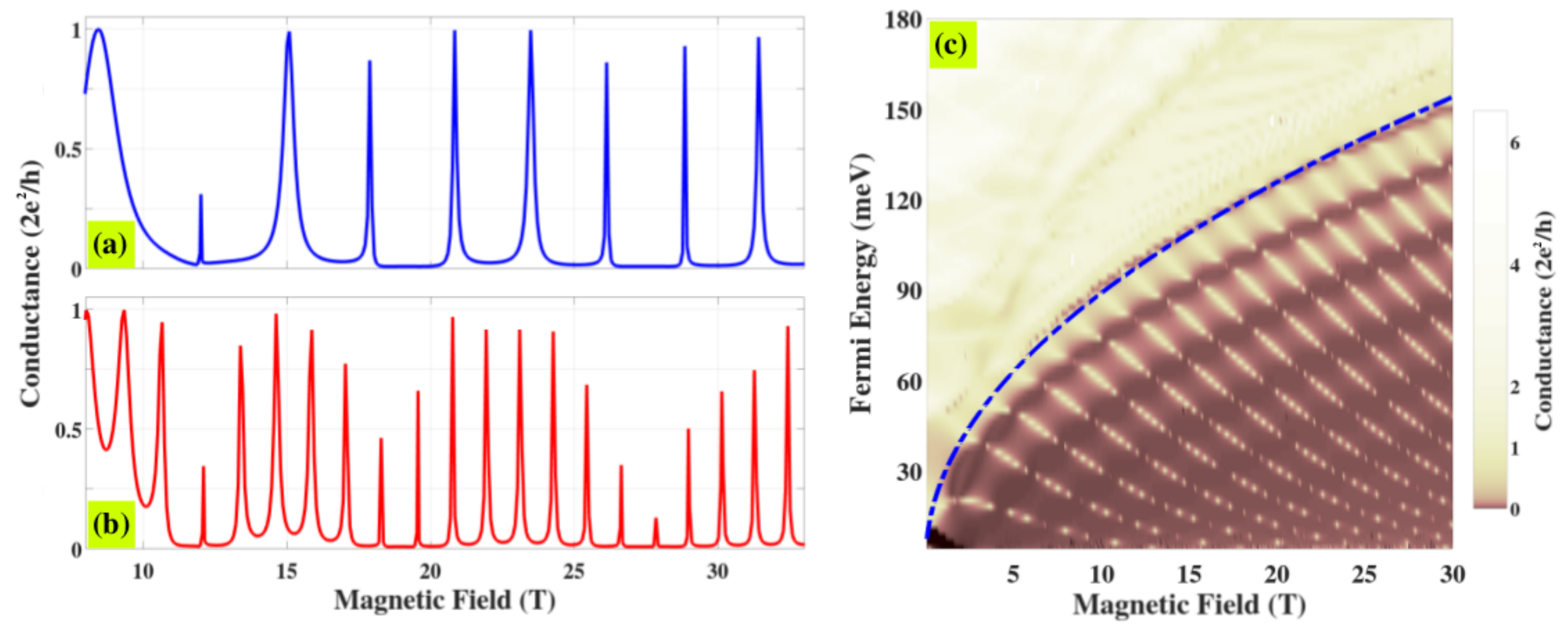}
	\caption{(a,b) Conductance as a function of \textit{B-}field in graphene nanoribbons containing two parallel defect lines separated by two different distances: $L_D \simeq 44$ nm and 88 nm, respectively. The ribbon width considered here is W $\simeq$ 50 nm and the Fermi energy $E_F = 80$ meV. (c) Conductance map with respect to \textit{B-}field and Fermi energy for $L_D \simeq 44$ nm. The superimposed blue-dashed line indicates the first Landau level $E_1 (B) = \sqrt{2e\hbar v_F^2 B}$ formed in the graphene grains.}
	\label{fig_sim3}
\end{figure*}
Firstly, it is necessary to distinguish the electron propagation around a grain boundary under strong \textit{B-}fields and that obtained in graphene \textit{p-n} \cite{Rickhaus2015} and \textit{B}-field heterostructures \cite{Oroszlany2008}. Due to the presence of opposite charge carriers in graphene \textit{p-n} junctions or opposite \textit{B-}fields in \textit{B}-field heterostructures, opposite cyclotron orbits are observed in the two sides of these systems, thus forming snake trajectories for electrons when they propagate along the system interface as illustrated in Fig.\ref{fig_sim1}.b. These snake trajectories consequently result in magneto-resistance oscillations \cite{Rickhaus2015}, provided that ballistic transport is present. On the contrary, around a grain boundary, a dramatically different picture is obtained (see Fig.\ref{fig_sim1}.c and the further demonstration in Sec. 1 of the Supplementary Information \cite{Nguyen_suppl}). In particular, because of scatterings at the grain boundary, electrons (following cyclotron orbits) are partly reflected and partly transmitted across the extended line of defects. While reflected electrons on the left side continuously propagate along the boundary to the bottom edge, transmitted electrons follow the similar cyclotron orbits on the right side and hence propagate in the opposite direction to the top edge (opposite skipping orbits - see Fig.\ref{fig_sim1}.c). Thus, two opposite propagation directions of electrons are predicted along the two sides of the grain boundary. This phenomenon differs completely from the picture in graphene \textit{p-n} and \textit{B-}field heterostructures, where electrons propagate in the same direction on both sides of the interface.

Despite such difference, electrons similarly follow the cyclotron orbits along the system interface and hence one can anticipate some similar features in the magneto-conductance in both cases. Indeed, the skipping trajectories illustrated in Fig.\ref{fig_sim1}.c can result in magneto-conductance oscillations in single grain boundary graphene systems (see the complete demonstration in the Supplementary Information \cite{Nguyen_suppl}), similar to those induced by the snake trajectories in graphene \textit{p-n} and \textit{B}-field heterostructures. However, note that this fascinating oscillation also requires electrons to transmit ballistically along the grain boundaries \cite{Nguyen_prepa}, i.e., it could be significantly disturbed by the scatterings along the disordered grain boundaries. Most importantly, the picture discussed above demonstrates that defect scatterings can be exploited to create the tunneling paths between quantum Hall edge states and hence AB interferometers can be obtained using graphene systems with two grain boundaries as proposed in Fig.\ref{fig_sim1}.

Fig.\ref{fig_sim2} presents the prediction of magneto-transport through a general system of polycrystalline graphene \cite{Qyu2011,Yasaei2014}, which indeed displays strong AB oscillations in the conductance (see Fig.\ref{fig_sim2}.a) at high magnetic fields. The atomistic model presents a similar geometry as in Fig.\ref{fig_sim1}.a, additionally, graphene grains exhibit different crystallographic orientations and disordered grain boundaries are considered (see its atomic structure described in detail in the Supplementary Information \cite{Nguyen_suppl}). The observed AB oscillations can be essentially explained as follows. First, under a strong magnetic field, electrons in the internal graphene grain have to propagate following quantum Hall edge states. Second, because of defect scatterings at the grain boundaries, these electrons can circulate inside such graphene grain (i.e., along its edge). Therefore, depending on the accumulated phase of electron wave in such circulating trajectories, constructive/destructive states are created and tuned when varying the \textit{B-}field, thus modulating the electronic conduction of the system as illustrated in Fig.\ref{fig_sim2}.a. To further visualize such picture, the local density of left-injected electronic states reflecting the left-to-right propagation of electrons is computed and displayed in Figs.\ref{fig_sim2}.b-c. The circulating trajectories mentioned above for electrons in the internal grain are indeed visibly observed. Simultaneously, strong backscatterings occur in the destructive state (see Fig.\ref{fig_sim2}.b), explaining the obtained low conductance, whereas the electron wave highly transmits through the system in the constructive one (see Fig.\ref{fig_sim2}.c).

Fig.\ref{fig_sim2} also demonstrates that strong AB interferences can be obtained in polycrystalline graphene even if there is a misorientation between grains and they present irregular edges. Together with the investigation on edge disorders and disorders at grain boundaries that will be presented below, this result demonstrates that the AB effect is quite robust under all these structural issues.

\begin{figure}[!t]
	\centering
	\includegraphics[width=3.4in]{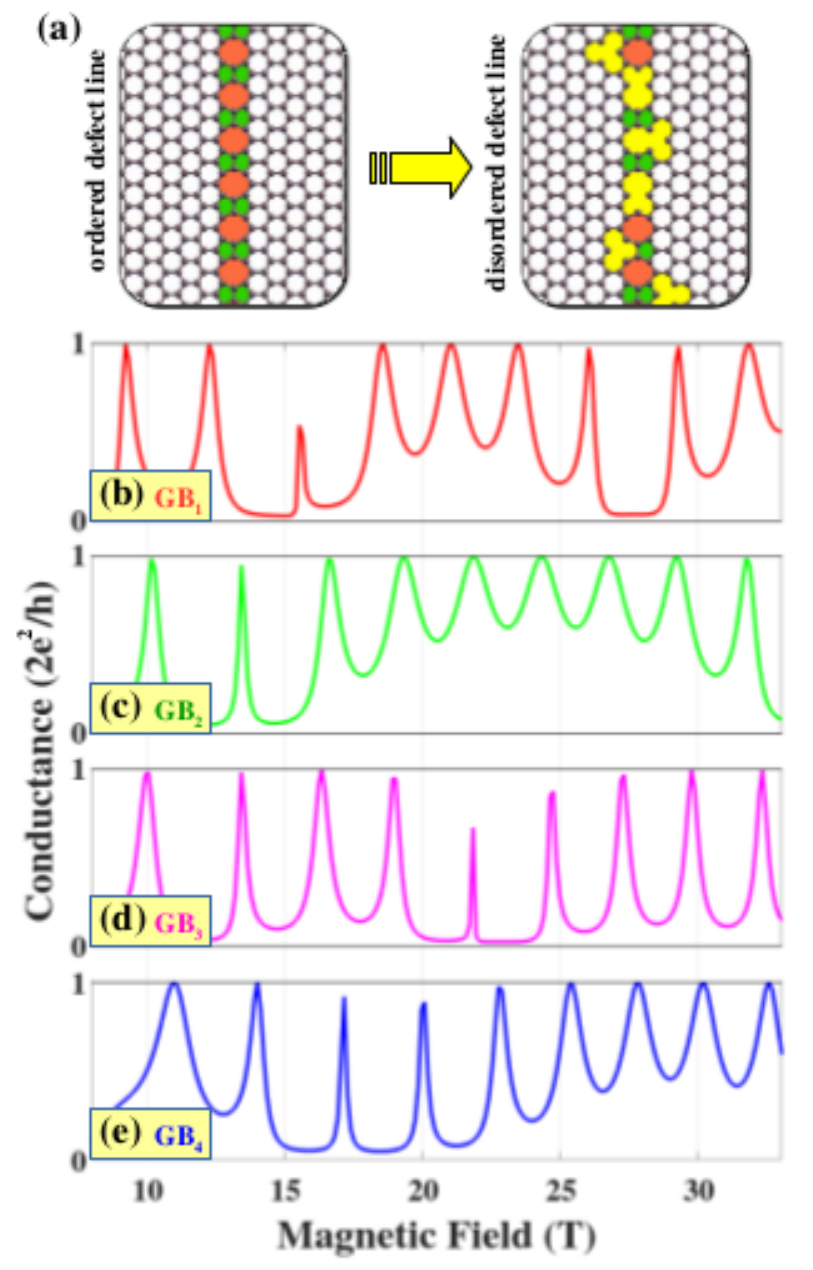}
	\caption{Aharonov-Bohm oscillations (b-e) in graphene nanoribbons containing different disordered defect lines (see (a)). The ribbon width W $\simeq$ 50 nm, distance between two defect lines $L_D \simeq 44$ nm and Fermi energy $E_F = 80$ meV are considered. In all presented configurations (from GB$_1$ to GB$_4$), $4 \%$ vacancies are randomly generated along/around the ordered defect line as illustrated in (a).}
	\label{fig_sim4}
\end{figure}

In order to discuss in detail the main properties of this AB interference, our investigation is now switched to conventional graphene ribbons containing two parallel defect lines \cite{Lahiri2010,Chen2014} (see the atomic structure in Fig.\ref{fig_sim4}.a), i.e., the misorientation between grains and the irregularity of system edges are neglected. The dependence of AB oscillations on the area of the internal graphene grain (particularly, the distance $L_D$ between defect lines) is displayed in Figs.\ref{fig_sim3}.a-b while Fig.\ref{fig_sim3}.c presents the corresponding conductance as a function of both magnetic field and Fermi energy. First, the period of these AB oscillations is shown to satisfy very well the standard formula $\Delta B = h/eS$ where $S$ is determined as the surface area enclosed by circulating channels inside the internal graphene grain. Note here that since these channels are obtained inside but not exactly at the edge of the internal grain as illustrated in Figs.\ref{fig_sim2}.b-c, the area $S$ is generally smaller than the area of such grain. In particular, the above formula accurately estimates the period of conductance oscillations in Figs.\ref{fig_sim3}.a-b using $S \simeq$ 1750 nm$^2$ (for $L_D \simeq 44$ nm) and 3490 nm$^2$ ($L_D \simeq 88$ nm), which are about $20 \%$ smaller than the values 2200 nm$^2$ and 4400 nm$^2$ of the area of the internal grain, respectively. 

Second, the results in Fig.\ref{fig_sim3}.c represent two distinguished energy regimes where strong AB oscillations are observed for low energies while they are blurred in the high energy regime. Actually, the superimposed dashed line separating these two regimes indicates the first Landau level formed in graphene grains (i.e., for the large graphene nanoribbons considered here): $E_1 = \sqrt{2e\hbar v_F^2 B}$ \cite{Yin2017} with the Fermi velocity $v_F = 3t_0 r_0/2\hbar$. Similar to analogous features reported in other systems \cite{Kolasinska2016,Nguyen2019,Nguyen2013b,Zhang2017}, they can be essentially explained by an inherent property of AB interference: strong conductance oscillations are achieved in the regime of single energy band, otherwise the effect can be disturbed by the multi-bands contribution. However, in contrast to the similar picture that has recently been reported in ref. \cite{Nguyen2019} for \textit{B-}field heterostructures, the intensity of the conductance peaks in the low energy regime still varies, depending on the carrier energy. This could be basically understood as a consequence of the nature of defect scatterings in the considered systems. At last, both results in Figs.\ref{fig_sim3}.a-b and \ref{fig_sim3}.c suggest that strong AB oscillations could, in principle, be measured experimentally at reasonably low magnetic fields (e.g., a few teslas) using large graphene grains and/or low carrier densities.
\begin{figure}[!t]
	\centering
	\includegraphics[width=3.4in]{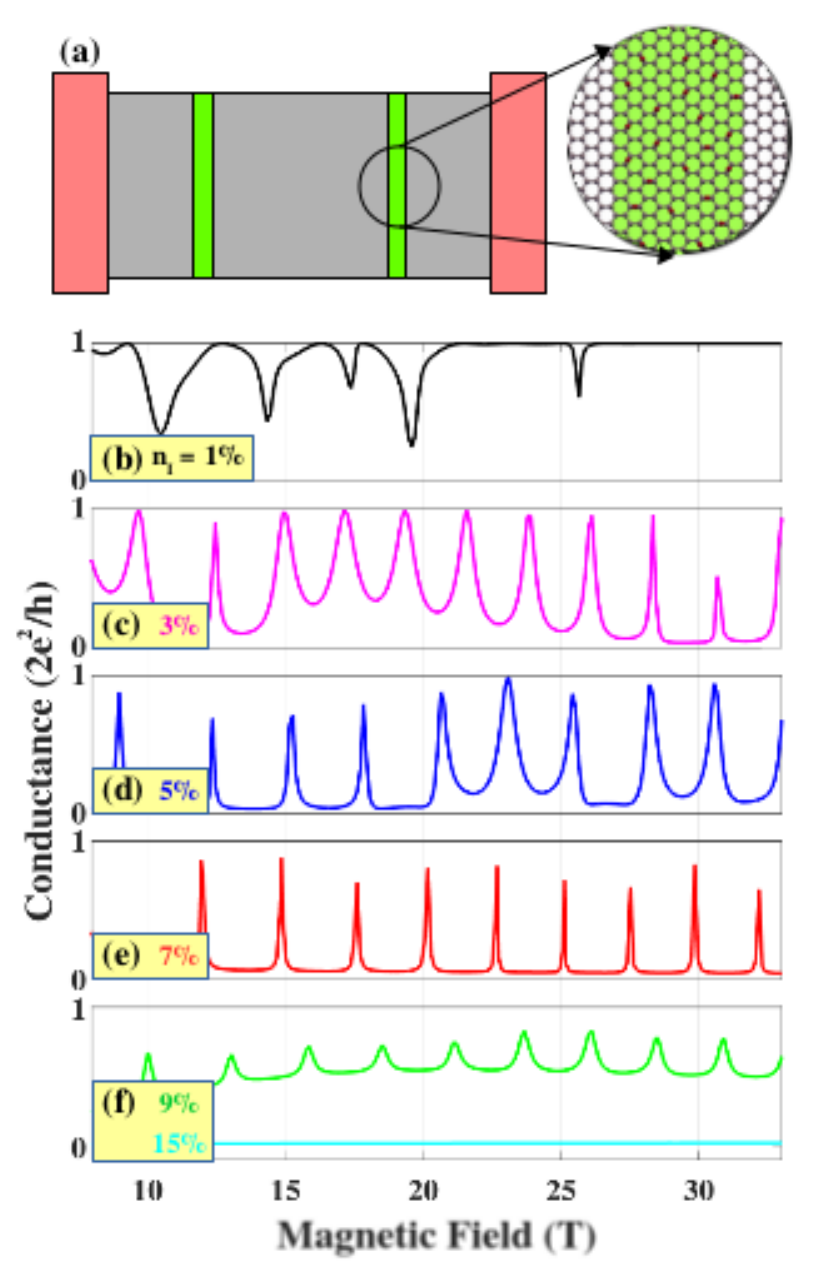}
	\caption{Aharonov-Bohm oscillations (b-f) in graphene nanoribbons containing two oxygen impurity barriers (see (a)) with different impurity densities $n_I$. The ribbon width W $\simeq 50$ nm, width of impurity barriers $L_{OI} \simeq 10$ nm, distance between two barriers $L_B \simeq 44$ nm, and Fermi energy $E_F = 80$ meV are considered.}
	\label{fig_sim5}
\end{figure}

Next, using these polycrystalline graphene systems naturally gives rise to a question related to the intrinsic effect of structural disorders. Two important disorder sources are considered: edge disorders and disorders at grain boundaries, which have been shown to be practically unavoidable. Similarly to recent results reported in ref. \cite{Nguyen2019} and as investigated in Fig.\ref{fig_sim2} for a system with irregular edges, strong conductance oscillations can still be achieved when taking into account the edge disorders, due to the robustness of quantum Hall edge states under the scatterings of such disorders (not further presented). Regarding the disorders at grain boundaries, many complex carbon rings, which are mostly formed by adding vacancies to ordered defect lines (containing hexagonal, pentagon, heptagon, and octagon membered-rings), can randomly occur. To estimate the effects of this disorder type, disordered defect lines were constructed by randomly introducing an amount of vacancies to / around the ordered one as illustrated in Fig.\ref{fig_sim4}.a. Magneto-transport obtained in four differently disordered systems (see more cases in the Supplementary Information \cite{Nguyen_suppl}) is presented in Figs.\ref{fig_sim4}.b-e. Indeed, these disorders induce modifications on the conductance oscillation, i.e., conductance peaks and valleys are changed depending on the disorder configuration (GB$_1$-GB$_4$), compared to those obtained in the ordered one (see Fig.\ref{fig_sim3}.a). Interestingly and similarly to the edge disorder cases, strong AB oscillations can still be achieved with reasonable vacancy densities (i.e., in the range $\leq 6 \%$ as presented in the Supplementary Information \cite{Nguyen_suppl} and up to about 10 $\%$ (not shown)). This robustness of AB oscillations can simply be understood as a result of the robustness of quantum Hall edge states, as already discussed for the edge disorders. Thus, it is worth emphasizing that this is certainly an advantage of AB interference in the quantum Hall regime as the disorders at grain boundaries are practically difficult to avoid and can easily destroy other fascinating phenomena such as valley dependent transport and electron optics \cite{Gunlycke2010,Nguyen2016b} in polycrystalline graphene.

At last, beside the structural defective systems, the present approach can be similarly applied in graphene structures with narrow barriers containing chemical impurities. Indeed, similarly to structural defects, chemical impurities with a reasonable concentration can also induce scatterings on the charge transport (see Fig.S6 in the Supplementary Information \cite{Nguyen_suppl}). Experimentally, several techniques (e.g., atomic force microscope lithography, solution-phase oxidation, photoexcited charge transfer, ...) to create graphene oxide barriers in graphene have been demonstrated \cite{Xwu2008,Byun2011,Masubuchi2011,Wang2012,Lee2016a,Lee2016b,Dago2018}. Remarkably, these barriers whose widths are down to sub-20 nm and tunable oxidation level (i.e., oxygen impurity density) have been successfully achieved. The impurity density can be further tuned using thermal, chemical and/or electrochemical reduction methods \cite{Mao2012,Mathkar2012,Toh2014,Guex2017,Oliveira2018}, for instance, as performed in ref. \cite{Wang2012}. In Fig.\ref{fig_sim5}, magneto-transport through graphene ribbons containing two narrow (10 nm width) oxygen impurity barriers is presented with different impurity densities $n_I$ (O/C ratios). Once more, strong AB oscillations are predicted with reasonable impurity densities (from 3$\%$ to 7$\%$ as seen) and are essentially due to the same mechanism as in the case of structural defects presented above. Note however that these oscillations are difficult to observe / can be destroyed by a large impurity density ($n_I$ exceeds about 10$\%$) and/or a large barrier width (see in the Supplementary Information \cite{Nguyen_suppl}, especially for the case of $n_I = 7\%$). This can be basically explained, similarly as in ref. \cite{Huang2012}, by the fact that when increasing the impurity density, a metallic-semiconducting transition appears and hence graphene barriers with oxygen impurities become wide-gap semiconductors, thus suppressing the charge transport as calculated in Fig.\ref{fig_sim5} for $n_I = 15\%$. Similar effects are observed when increasing the barrier width because impurity scatterings consequently increase and can suppress the transport quantities when they are strong enough. A rough estimation for oxygen impurity barriers with impurity density in the range from 2$\%$ to 8$\%$ and, accordingly, barrier width of about 5 nm to 25 nm and could be used to experimentally achieve significant AB oscillations.

Finally, it is very worth noting that since it is demonstrated without taking into account the bipolar characteristics of graphene, the present approach can, in principle, be also applied to similar systems of 2D materials beyond graphene and their lateral heterostructures \cite{Lin2016,Ovando2019}, thus representing a large range of possibilities to design these AB interferometers.

\section{Conclusion}

In summary, magnetotransport through polycrystalline graphene systems was investigated using atomistic tight-binding calculations. These quantum simulations demonstrated that defect scatterings can be exploited to create the transmission between quantum Hall edge states. Consequently, strong Aharonov-Bohm oscillations can be achieved in realistic systems containing two parallel grain boundaries when strong magnetic fields are applied. The properties and dependence of these Aharonov-Bohm oscillations on parameters such as charge carrier energies, graphene grain sizes and structural disorders were clarified. In addition, similar Aharonov-Bohm interferences were also predicted in graphene nanoribbons presenting two separated barriers which contain randomly distributed chemical impurities. Finally, since this phenomenon is achieved in the unipolar regime, the present approach can be perspectively applied to other 2D layered materials beyond graphene.   

%\section*{Conflicts of interest}
%There are no conflicts to declare.

%\section*{Acknowledgments}
%The authors acknowledge financial support from the F.R.S.-FNRS of Belgium through the research project (N$^\circ$ T.1077.15), from the Flag-Era JTC 2017 project "MECHANIC" (N$^\circ$ R.50.07.18.F), from the F\'{e}d\'{e}ration Wallonie-Bruxelles through the ARC on 3D nanoarchitecturing of 2D crystals (N$^\circ$ 16/21-077) and from the European Union's Horizon 2020 research and innovation program (N$^\circ$ 696656).

\textbf{Acknowledgments} - The authors acknowledge financial support from the F.R.S.-FNRS of Belgium through the research project (N$^\circ$ T.1077.15), from the Flag-Era JTC 2017 project "MECHANIC" (N$^\circ$ R.50.07.18.F), from the F\'{e}d\'{e}ration Wallonie-Bruxelles through the ARC on 3D nanoarchitecturing of 2D crystals (N$^\circ$ 16/21-077) and from the European Union's Horizon 2020 research and innovation program (N$^\circ$ 696656).

\newpage

\onecolumngrid

\includegraphics[page=1,scale=0.9]{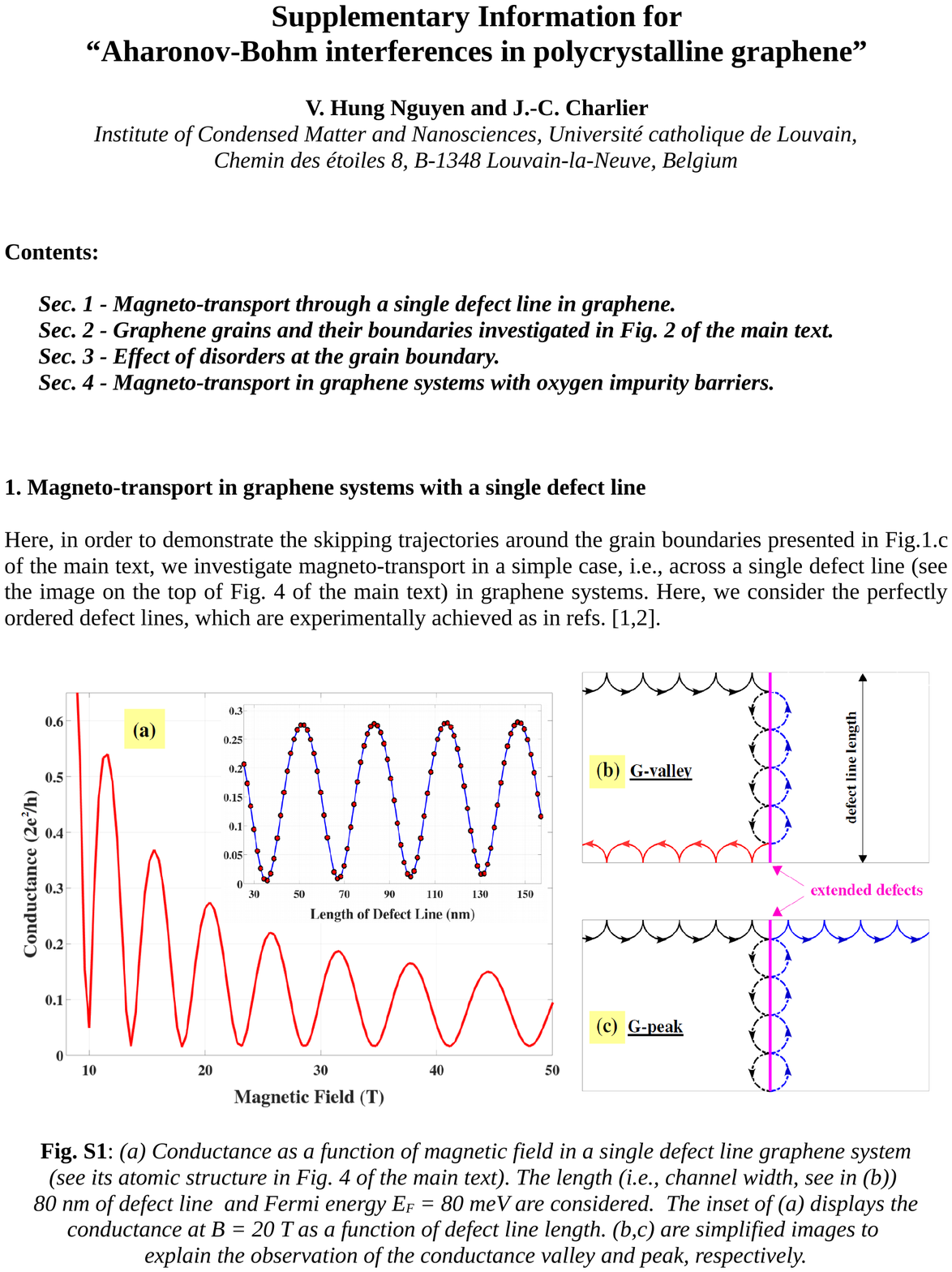} 
\includegraphics[page=2,scale=0.9]{Nguyen_suppl.pdf}  
\includegraphics[page=3,scale=0.9]{Nguyen_suppl.pdf}  
\includegraphics[page=4,scale=0.9]{Nguyen_suppl.pdf}  
\includegraphics[page=5,scale=0.9]{Nguyen_suppl.pdf}  
\includegraphics[page=6,scale=0.9]{Nguyen_suppl.pdf}  
\includegraphics[page=7,scale=0.9]{Nguyen_suppl.pdf}  

\end{document}